\documentclass{aa}  
\usepackage{graphicx}
\usepackage{epsfig}

\usepackage{txfonts}
\begin{document}
%
\title{Toward a Clean Sample of Ultra-Luminous X-ray Sources}
   \subtitle{}
   \author{M. L\'opez-Corredoira\inst{1}, C. M. Guti\'errez\inst{1}}
   \offprints{martinlc@iac.es}

\institute{
$^1$ Instituto de Astrof\'\i sica de Canarias, C/.V\'\i a L\'actea, s/n,
E-38200 La Laguna (S/C de Tenerife), Spain}

   \date{Received xxxx; accepted xxxx}

  \abstract
   {Observational follow-up programmes for the characterization of ultra-luminous X-ray
   sources (ULXs) require the construction of clean samples of such sources in which the
   contamination by foreground/background sources is minimum.}
   {In this article we calculate the degree of foreground/background 
   contaminants among the ULX sample candidates in the Colbert \& Ptak 
   (2002) catalogue and compare these computations with available 
   spectroscopical identifications.}
   {We use statistics based on known densities of
   X-ray sources and AGN/QSOs selected in the optical. The analysis is done 
   individually for each parent galaxy. 
   The existing identifications of the optical counterparts are compiled
    from the literature.}
   {More than a half of the ULXs, within twice the distance of
   the major axis of the 25 mag/arcsec$^2$ isophote from 
   RC3 nearby galaxies and with X-ray luminosities $L_X$[2-10 keV] $\ge 10^{39}$ 
   erg/s, are expected to be high redshift background QSOs.
   A list of 25 objects (clean sample)  confirmed
   to be real ULXs or to have a low probability of being contaminant 
   foreground/background objects is provided.}
   {}
   
\titlerunning{Toward a Clean Sample of ULXs}
\authorrunning{L\'opez-Corredoira \& Guti\'errez}
   \maketitle
%

\section{Introduction}

ULXs (ultraluminous X-ray sources) are objects associated with some nearby
galaxies, and  which emit in X-ray with luminosities above the Eddington limit 
(so they cannot be stellar-mass X-ray binaries), 
but which are well below the typical emission in the center of AGNs. 
The physical nature of ULXs still is unclear.
One of the most interesting possibilities is that they are black holes of 
intermediate masses (IMBHs; Colbert \& Mushotzky 1999;
Makishima et al. 2000).
Follow-up programmes over wide spectral ranges are essential for
revealing the nature of these sources. 

There is strong observational evidence
that points to the existence of other kinds of objects linked to the parent
galaxies: 
short-lived massive stars, usually in HII regions (e.g.\ Pakull \&
Mirioni 2002; Gao et al.\ 2003; Kaaret et al.\ 2004; Zampieri et al.\ 2004; Liu et
al.\ 2004; Smith et al.\ 2005; Lehmann et al.\ 2005), tentatively 
associated with intermediate-mass black holes. 
The ULX phenomenon is indeed  closely connected
to star formacion activities, since ULXs preferentially occur in 
spiral arms of late-type or starbursts
rather than in early-type galaxies (Kilgard et al. 2002;
Schwartz et al.\ 2004; Liu et al.\ 2005b). Actually, 70\% of starburst 
galaxies host at least one ULX. However, as has been shown in
the identification of several optical counterparts of ULXs (Lira et al.
2000; Arp et al.\ 2004; 
Guti\'errez \& L\'opez-Corredoira 2005, 2006; Guti\'errez 2006; Wong et al.\
2005; Clark et al. 2005; e.g., IXO 1, 2 in NGC 720; 
see Table \ref{Tab:main}) the degree of contamination by foreground/background 
objects is very important. 
In any case, the contamination is not total, so
it is incorrect to think that all ULXs are QSOs with different
redshifts (Burbidge et al.\ 2003).

Several statistical analyses (e.g.\ Ptak \& Colbert 2004; Schwartz et al.\ 2004;
Irwin et al.\ 2004) of current compilations of ULXs confirm the relevance of such
contamination. 
In this paper we address the problem of such a
contamination in existing samples of ULXs, extending the work by these authors. 
We analyse the catalogue
by Colbert \& Ptak (2002) of ULX candidates and estimate the contamination in
each galaxy and  the most likely truly ULX sources. The paper is organized as
follows: Section~2 describes the catalogue used and the corrections applied to obtain
an homogeneus sample; the distribution of background sources and the method of
computing probabilities are  considered in Sections~3 and 4; Section~5 presents
the results of the statistical analysis; and Sections~6 and 7 present the discussion
and conclusions respectively, 
analysing also the possibility of some 
non-cosmological redshifts.


\section{Catalogues of ULXs and sample selection}

There are several catalogues of ULXs (ultraluminous X-ray sources), also called
IXOs (intermediate-luminosity X-ray objects), 
built from ROSAT, XMM or Chandra data.
Colbert \& Ptak (2002) (C02 hereafter), and Liu \& Bregman (2005) are based in
ROSAT data, while Schwartz et al. (2004) used Chandra data, and Liu \& Mirabel
(2005) did a compilation of the literature with a mixture of different data from
different surveys, including data from Chandra and XMM X-ray telescopes. We
decided to analyse the C02 catalogue because i) an important fraction of the
sources have been identified in the optical, and this allows a direct comparison
with our statistical results; ii) the objects have been selected using data
from just one X-ray instrument, doing a homogeneous sample; 
and iii) a correction is possible that
includes objects that were known to be background objects.

The analysis should not be carried out on the sample as a whole because we are
dealing with a catalogue with constraints in luminosity rather than in flux for
galaxies within a wide range of distances; we cannot do a simple calculation
such as multiplying the total covered area by the average density of QSOs up to
a given magnitude, because the limiting magnitude is different for each galaxy.  
Analysing the galaxies individually
will also enable us to identify  the galaxies  likely to be
totally/partially contaminated and those that are not,
and will provide a useful tool for ULX searches and  optimizing future follow-up studies. 
Furthermore, we  compile the recent literature about
the observation of the optical counterparts of these candidate ULXs that will
corroborate our statistical analysis or will show whether there is some 
possible anomalous excess of different redshift objects around the parent
galaxies.

C02 presented a catalogue of 87 candidate ULXs, 
which were observed in X-ray  with
ROSAT HRI  (Voges et al.\ 1999); they were around bright galaxies from the
RC3 catalog whose velocity is less than 5000 km/s and have more than 90\% 
of the galaxy in the field of view of the ROSAT HRI observations (the total number
of galaxies within these constraints was 766). They were detected by simply
measuring the luminosity of these ROSAT sources assuming they are at the distance
of a parent galaxy that is nearby: their sample of IXOs follows the constraint
that the ULX is at an angular distance less than 2$R_{25}$ from the centre of an RC3
galaxy, where $R_{25}$ is the major axis of the   25 mag/arcsec$^2$ isophote,
and the luminosity assuming the ULX has the same distance as the galaxy
$L_X[2-10 keV]\ge 10^{39}$ erg/s. A total of 54 galaxies had at least one of
the 87 candidate ULXs with these constraints, and some galaxies were surrounded by
up to seven candidate ULXs.

C02 have not included in their catalogue some sources observed by ROSAT HRI that
could be classified as candidate ULXs according to their constraints because
they were known to be QSOs with a redshift different from that of the RC3 galaxy. 
Arp et al.\ (2004, hereafter A04) (table 1) list these sources; there are 12 extra
candidate ULXs in this list which follow the C02 constraints.  Even if they are
background sources that have been discarded as ULXs associated with the galaxy,
they  should be included in the catalogue if we wish to select a statistical
sample.  Indeed, as we will see throughout this paper, many of the candidate ULXs
turned out later also to be background sources, so there is no reason to separate
sources that were known to be background sources before or after the C02 compilation. 
If we include these sources, there are 99 candidate ULXs around 57 galaxies. Hereafter
we will denote this sample as C02c ('Colbert \& Ptak 2002 corrected'). 
They are listed in Table \ref{Tab:main}.



\section{Distribution of background sources}

To estimate the contamination in the C02c catalogue
we  consider the distribution of  background sources in two
different wavelength ranges: i) any source,
$W\equiv X$, X-ray at frequencies between
0.5 and 2.0 keV; ii) QSOs, $W\equiv B$, B optical filter.

For the X-ray, we use the cumulative counts by Hasinger et al.\ (1993) obtained
in the Lockman Hole from ROSAT-PSPC counts using the flux 
spectral-shape they also derived. 
A parameterization of their fig.\ 8 with two linear
regimes gives (in units of deg$^{-2}$) as a function of the flux $S_{X,i}$ (units:
erg/s/cm$^2$):

\[
N_{X,i}(>S_i)=
\]\begin{equation}
=\left \{ \begin{array}{ll}
        1.0\times 10^{-21}S_{X,i}^{-1.66},& \mbox{ $S_{X,i}\ge 2\times 10^{-14}$} \\

        2.8\times 10^{-12}S_{X,i}^{-0.97},& \mbox{ $S_{X,i}<2\times 10^{-14}$}
\end{array}
\right \}
\label{NX}
.\end{equation}

This law is valid in principle for the range $10^{-15}<S_{X,i}<
10^{-12}$, where most of our ROSAT sources are.
Hasinger (1998) gives the counts for a wider regime, and 
the linear regimes are present in
wider ranges, although the exponents change slightly for larger values
of $S_{X,i}$. Zickgraf et al.\ (2003, fig. 18) give the counts for
brighter X-ray sources.
In any case, for the estimation of probabilities (see next section),
the law of eq.\ (\ref{NX}) is accurate enough.

We take the ROSAT X-ray flux of the sources, $S_{X,i}$, from C02c. 
They give the luminosity for the range 2 KeV$<\nu
<$10 KeV but this can be converted into fluxes if we divide the luminosity by
$4\pi d_{galaxy}^2$; and $S_{X,i}\equiv S_i[0.5\ KeV<\nu <2\ KeV$] = 0.548
$S_i[2\ KeV\nu <10\ KeV]$. The factor 0.548 stems from the assumption
that the X-ray spectrum of the sources is
$S[\nu ]d\nu \propto \nu ^{-0.7}d\nu $, an assumption 
 also adopted by C02 to calculate their luminosities.

For the estimation of the background QSOs in the B-band, 
we use the counts by Cheney \& Rowan-Robinson (1981) for $m_B\le 18.5$
and Boyle et al. (2000) for $m_B>18.5$ (see the fit in fig.\ A.1
of L\'opez-Corredoira \& Guti\'errez 2004) as a function of the photographic
magnitude $m_{b_j,i}$. Where no B-magnitude is available for a 
candidate ULX optical counterpart, we use the
V-magnitude (in QSOs, $(B-V)$ is small). Since we are only interested in a 
rough approximation of the counts, no colour correction for the difference
between photographic and Johnson filter or with V in case of few QSOs from
A04 is applied. Boyle et al.\ (2000) counts of colour-preselected
QSOs agree more or less 
with the counts from a more complete spectroscopic survey by Meyer et al.\ (2001).
The counts include both QSOs and Seyfert 1 galaxies (Seyfert 1-type galaxies would form $\sim $10\% of
the total counts) but not Seyfert 2, LINERs or other types of AGNs. According
to Hao et al.\ (2005), Seyfert 2 would be less than four time more numerous
than Seyfert 1, so the inclusion of Seyfert 2 would increase the counts
up to  40\% . Most of our sources are presumably point-like QSOs rather 
than extended galaxies, given the images of the optical counterparts.
In any case, since we are interested in the order of
magnitude, it is not so important whether an AGN is a QSO or a Seyfert
galaxy. 

The counts are (units deg$^{-2}$):

\begin{equation}
N_{B,i}(<m_{B,i})=
\left \{ \begin{array}{ll}
        10^{-2.8+0.8(m_{B,i}-15)},& \mbox{ $m_{B,i}\le 18.5$} \\

        1981.0-214.2m_{b_j,i}+5.792m_{b_j,i}^2,& \mbox{ $m_{B,i}> 18.5$}
\end{array}
\right \}
\label{NB}
.\end{equation}

In this case, the B magnitudes were taken from the
 NOMAD database\footnote{http://www.navy.mil/nomad.html}
of the most likely optical counterpart (the brightest
candidate closest to and within 10$''$ of the X-ray source). 
The nominal positional accuracy of ROSAT HRI is 5 arcseconds
(Voges et al.\ 1999). 
If we compare the position of the 14 candidate ULXs common to CP02
(with ROSAT) and Schwartz et al.\ (2004) (with Chandra), we find
that the average difference is 2.4 arcseconds.
If we do not find any counterpart within 10$''$, 
we attribute this a magnitude $m_B>21.5$,
i.e.\ higher than the limiting magnitude of the Palomar plates.
It may happen that the optical source is the false counterpart
because the real ULX is much fainter; 
however, the probability is not very
high in regions away from the galaxy. In fact, the probability
of observing an object with magnitude $\sim 20$ within a circle
of radius 10$''$ is $\le 10^{-3}$ for QSOs (Boyle et al.\ 2000)
and $\sim 10^{-2}$ for high galactic latitude stars (Robin et al.\ 2003),
which we consider negligible.
A higher contamination could be present in spiral arms, where
the density of new-born stars over HII regions could be high. 
For instance, Kaaret (2005)
shows that IXO 5 is much fainter than the magnitude of the brightest
source in A04, found as an optical counterpart. 
In this case, and when spectroscopy
of the optical counterpart has not revealed a QSO or stellar nature,
the contamination with regard the visible magnitude 
would be underestimated but never overestimated, so we
will get a {\it minimum} of the ratio of contamination.

\section{Probabilities of being background sources}

From the catalogue of ULXs in Table \ref{Tab:main} we want to calculate
the probabilities that they are background objects (mainly background QSOs) 
instead of X-ray sources associated with their putative parent galaxies.

Given a galaxy with $n$ ULXs around whose angular distances from its centre
are $\alpha _i$ (deg) and background density $N_{W, i}$ (deg$^{-2}$) 
of sources brighter than the flux/``apparent magnitude'' of this source in 
a wavelength range $W$, the probability that all of them are background 
sources is:

\begin{equation}
P_W\sim e^{-\lambda _{W:max}}\left(\sum _{i=n}^\infty 
\frac{\lambda _{W:max}^i}{i!}\right)
\left(\prod _{j=1,j\ne i_{W:max}}^n
\frac{2\lambda _{W,j}}{\lambda _{W:max}}\right)
\label{P}
,\end{equation}
\[\lambda _{W:max}=Maximum[\lambda _{W,i}]; \ \ \ \ 
i_{W:max} \ \ / \ \ \lambda_{W,i_{max}}=\lambda _{W:max}\]
\[
\lambda _{W,i}=\pi \alpha _i^2N_{W,i}
\]
The  first two factors represent a Poissonian distribution for
 $n$ background sources equivalent to $i_{W:max}$ 
(the fainter and more distant source). The third 
factor, $\left(\prod _{j=1,j\ne i_{W:max}}^n
{2\lambda _{W,j}}/{\lambda _{W:max}}\right)$,
takes into account that the sources might be 
brighter or closer to the centre of the galaxy than normally expected
in a random process. Randomly, one would expect that 
${\lambda _j}/{\lambda _{max}}\approx {1}/{2}$ on average, so this
third factor should be nearly unity; but in some cases, some sources are
much brighter than $i_{W:max}$ or much closer to the centre than $i_{W:max}$,
and in such cases the third factor would be much lower than one,
reflecting the fact of this improbable difference in brightness or angular
distance. For example, in the case with three candidate ULXs with magnitudes
20, 16 and 15 respectively and distances around 1 arcminutes 
from the nuclei of the galaxy, we will calculate the Poissonian probability
of having background QSOs within a circle of 1 arcminute with magnitude lower
than 20 (with the first two factors), and the third factor will reduce this probability
to reflect the fact that the QSOs with magnitudes 15 and 16 are much brighter 
(improbable) than those with magnitudes 19--20. As another example,
imagine three QSOs of roughly the same magnitude 20 with distances to the
galaxy 0.1, 0.2 and 3.0 arcminutes respectively. We will calculate
the Poissonian probability
to have background QSOs within a circle of 3 arcminutes with magnitude 20
(with the first two factors), and the third factor will reduce this
probability to reflect that the QSOs with distances 0.1 and 0.2 arcminutes
are much closer than expected randomly.

Table \ref{Tab:main} gives $\lambda _{X,j}$ and $\lambda _{B,j}$ 
for each source, and the global probabilities $P_X$ and $P_B$ for
each galaxy. The X-ray fluxes and the B-band magnitudes are already
listed in this table. And, if there is an optical identification, this
is indicated in the column 9.

\section{Results}

\subsection{Overall contamination}

A rough calculation with X-ray probabilities 
will give the total number of background sources:

\begin{equation}
{\rm number\ of\ background\ sources}\approx \sum _{j=1}^{99}
(1-e^{-13.4\lambda _{X,i}})=50.3
,\end{equation}
where the factor $13.4={766}/{57}$ is the ratio of the
total number of galaxies to those with at least
one candidate ULX.
The total number of background sources that we obtain with this
expression is  51\% of the background sources.
It would be $>51.2$ ($>52$\%, only QSOs) if we did the calculation 
with $\lambda _{B,i}$ instead of $\lambda _{X,i}$.

So far, 54 sources have been already identified if we include the diffuse 
objects as identified (see Table \ref{Tab:main}), 
and 36 of them ($67\pm 11$\%) turned out to be 
contamination; 18 of the 
ULX candidates are indeed sources belonging to the parent galaxy.
This is slightly higher than our statistical estimate in X-ray. 
We must also
bear in mind that there may be a bias in the identified
sources. They are on average brighter in the visible than the unobserved 
sources, although in principle fainter objects should have an even 
higher probability in the visible to be contamination. Other
selection effects such as the avoidance of crowded regions
in the spiral arms near the centre of the galaxies,
where many optical counterparts are possible, could reduce this probability.

Ptak \& Colbert (2004) calculated a similar contamination: around 44\%.
Liu et al.\ (2005b) calculated a contamination ratio about a 
30\% within $2R_{25}$. Swartz et al. (2004) conclude that 
$\sim 25$\% of the sources ULXs candidates 
from Chandra ACIS may be background objects 
including 14\% of the ULX candidates in the sample of spiral galaxies and 
44\% of those in elliptical galaxies, but their numbers are not
directly comparable with that of the C02 sample because in a fair number
of cases the Chandra fields do not cover the circles within $2R_{25}$.
For Irwin et al. (2004) 
the contamination is nearly total for elliptical galaxies and for the 
sources far from the centre in spiral galaxies. 
As said in the introduction, other authors
(Kilgard et al. 2002; Schwartz et al.\ 2004; Liu et al.\ 2005b)
also predicted a lower contribution of background
sources for late-type galaxies.

\subsection{Contamination in each galaxy}

Table \ref{Tab:main} presents the results of the statistical analysis. The expected  mean
number of random sources within the angular area are given in columns 4 and 7
for the analyses in the X-ray and in the optical respectively. The
probabilities that all the ULXs detected in a given galaxy are contaminants are
given in columns 5 and 8. Since C02c had a sample of 766 galaxies from which
there are 57 galaxies with at least one candidate ULX, only galaxies with $P_X$
or $P_B$ lower than $7\times 10^{-5}$ will have within 95\% C.L. at least one
X-ray source belonging to the galaxy instead all of them of being  background
sources. Of the 57 galaxies, only 18 of them follow this:  NGC 1073, NGC
1313, NGC 1399, NGC 1672, Holmberg II, NGC 3628, NGC 4088, NGC 4319, NGC 4374,
NGC 4472, NGC 4485, NGC 4559, NGC 4631, NGC 4649, NGC 4861, NGC 5128, NGC 5204 and
NGC 5194. The average Hubble type of these galaxies is +3.2. 
We found some ellipticals (Hubble type: $-$5 or $-$4)
with a low probability of all their ULXs being background sources, 
although no ULX was confirmed spectroscopically in them. 
See \S \ref{.anom} for a discussion on their low probabilities of 
being background sources.

The second group
with the remaining 39 galaxies would have 53 candidate ULXs, which would be
mostly background sources: either QSOs or any other kind of X-ray emitter. The
average Hubble type of the galaxies of this second group is +2.0. 
Indeed 22
sources of this second group have already a spectroscopic identification and
19 of them are QSOs or emission line galaxies with different redshifts from that 
of the main galaxy (see Table \ref{Tab:main}). Another source is a foreground M
star. Moreover, if one looks at the photometry of the
fourth release of the SDSS (Adelman-McCarthy et al.\ 2006), 
one will find within 10$''$ optical counterparts of seven IXOs of this second group
(4, 44, 48, 63, 64, 74 and 83)
and three of them (4, 44 and 74) have colours typical  of QSOs [according to 
the criterion in eq.\ (1) of L\'opez-Corredoira \& Guti\'errez 2004, which means
approximately that they have a probability of being real QSOs of 2/3].

This statistical analysis implies that, if one looks for ULXs, 
one should preferably observe the candidate ULXs of the 18
galaxies from the first group (with $P_X$ or $P_B$ lower than $7\times
10^{-5}$).
For many of these 18 galaxies, spectroscopic identification is also available:
there are confirmed sources belonging to the main galaxy for
IXOs 5, 7, 8, 31, 65, 68, 72, 73, 77, 80, 81 (see Table \ref{Tab:main}). 
Some other candidate ULXs are known to be bright diffuse sources that 
cannot be background QSOs: IXOs 26, 27, 42, 62, 79.
And there are also some identified QSOs with redshifts different
from those of the parent galaxies: 
IXOs 19, 20, 57, 58, 69, 71; A3, 5, 9, 10, 13, 14, 21, 22.
IXO75 is a foreground local M star;
therefore, there remain still to be identified 15 IXOs of the first
group: IXOs 17, 18, 25, 39, 50, 51, 55, 56, 59, 60, 61, 66, 70,
76, 78. The variable X-ray source
IXO 76 has  a possible optical counterpart corresponding
to a late O or early B star with magnitude 24.1 in the F555W filter 
(Ghosh et al.\ 2005), but this has not been confirmed spectroscopically.
If one looks at the fourth release of the SDSS 
(Adelman-McCarthy et al.\ 2006), one will find
optical counterparts of IXOs 50, 51, 55, 56, 59, 60, 61 and 70,
among which IXOs 50, 51, 55, 59, 70 have colours typical  of QSOs [according to 
the criterion in eq.\ (1) of L\'opez-Corredoira \& Guti\'errez 2004; although
they might also be HII regions with the same colours], while
IXOs 56, 60, 61 have a different colour, so they should be ULXs, foreground
stars, or emission line galaxies.


\subsection{Statistics with the clean sample}

We define the clean sample as the objects that are confirmed
ULXs or candidates with low probabilities of being contaminants.

According to the discussion in the previous subsection, the 
objects that have not yet been observed spectroscopically,
and that have higher probabilities of being ULXs in the C02 catalogue are 
IXOs 17, 18, 25, 39, 66, 76 and 78.
We removed  
IXOs 50, 51, 55, 59 and 70 from the initial list for their typical  QSOs colours without
being diffuse nebulae.
We also removed IXOs 56, 60, 61 because their $\lambda _{B,j}>1$
while the $P_B$ and $P_X$ of the parent galaxy NGC 4472 are not so low 
as to forbid a projection of seven QSOs (or any other kind of 
contaminating object).

The objects that are spectroscopically confirmed ULXs are (see Table
\ref{Tab:main}) IXOs 5, 7, 8, 22, 31, 34, 65, 68, 72, 73, 77, 80, 81; 
plus IXOs 26, 27, 42, 62, 79, which were not observed spectroscopically
but  are bright diffuse sources, so they cannot be background QSOs,
and are most probably HII regions of the parent galaxies.

\begin{figure}
\begin{center}
\vspace{1cm}
\mbox{\epsfig{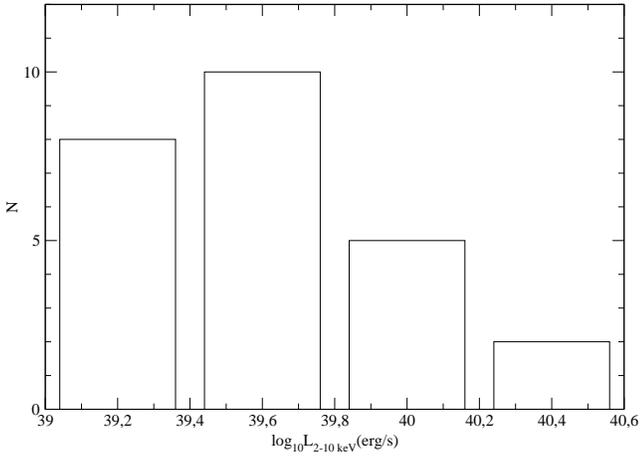}}
\end{center}
\caption{Distribution of luminosities in the clean sample of
ULXs.}
\label{Fig:histolum}
\end{figure}

\begin{figure}
\begin{center}
\vspace{1cm}
\mbox{\epsfig{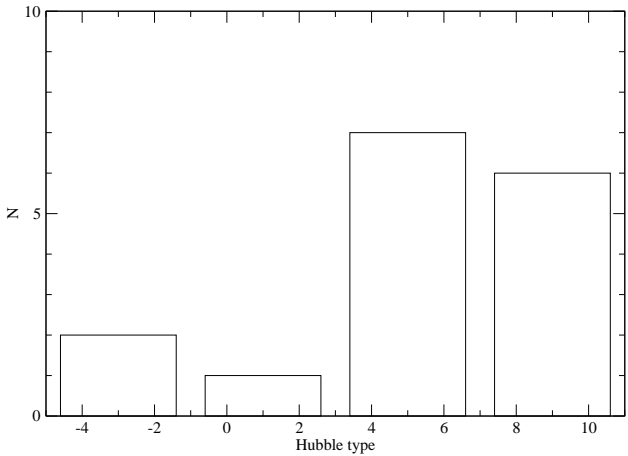}}
\end{center}
\caption{Distribution of Hubble types of the galaxies with
at least one member of the clean sample of ULXs.}
\label{Fig:histohub}
\end{figure}

In total, there are 25 objects. This sample might not be totally clean; 
even the objects identified as HII of the parent galaxy might correspond
to wrong optical counterparts,
but at least the contamination will be much lower than the original C02 sample.
We will repeat statistics made by Ptak \& Colbert (2004) on
C02 sample, but only with these 25 objects. A histogram of
energy distribution is shown in Figure \ref{Fig:histolum},
and a histogram of the distribution according to  Hubble type
is presented in Figure \ref{Fig:histohub}.

The distribution of energy peaks in intermediate energies 
($\sim 4\times 10^{39}$ erg/s), or could be flat for lower
energies (the Poissonian errors do not allow a distinction).
The distribution of the Hubble types peaks around late-type
spirals, as expected; however, we have still two ellipticals.
An examination of the IXOs from these ellipticals makes us
realize that they come from the unobserved candidates
around the elliptical galaxies NGC 1399 and NGC 5128. 
One is tempted to claim that
these candidate ULXs are indeed background QSOs, but in such
a case, we would have the problem of an anomalously high concentration of QSOs around 
NGC 1399 (see \S \ref{.anom}) and an inexplicably low value
of $P_X$ in NGC 5128 due to contamination.
NGC 1399 is indeed a cD galaxy
at the centre of a cluster; perhaps the ULXs near NGC 1399 belong to
nearby galaxies in the cluster.
The three ULXs in these two galaxies (IXOs 17, 18, 76) have not yet been observed;
 they are very faint (no optical counterpart has yet been found),
so we cannot confirm whether the expectation by Irwin et al.\ (2004) that
there are no ULXs associated with ellipticals is correct or not.

\section{QSOs with low probability to be background sources}
\label{.anom}

ULXs around some galaxies have been proposed (Burbidge et al.\ 2003) to be QSOs or
BL Lac objects with intrinsic  redshift being ejected from a parent galaxy 
rather than black holes with masses $10^2-10^4$ M$_\odot$. It is clear, as said
in the introduction, that this cannot be the case for all candidate ULXs because
there are many whose spectroscopy have shown a different nature. We
have tested whether the Burbidge et al.\ hypothesis could be applied to some of the
objects. Let us do a statistical analysis for each individual galaxy of the
first group (with $P_X$ or $P_B$ lower than $7\times 10^{-5}$) in which some
QSO has been observed:

\begin{description}

\item[NGC 1073]: with A3,5 as QSOs. Since IXO 5 is not a QSO,
the probability of being background objects is not so low for only two QSOs 
(excluding IXO 5 from the list): $P_X=10^{-3}$, $P_B=10^{-4}$; 
quite possibly background
sources. Note, however, that there is a third QSO around NGC 1073 which
was not included here because the X-ray flux is less than the C02 requirements
(equivalent to a luminosity $L_X[2-10 keV]\ge 10^{39}$ erg/s if they were
at the distance of the galaxy). With this third QSO, the probability
becomes lower, as claimed by Arp \& Sulentic (1979).

\item[NGC 1399]: with IXOs 19, 20 as QSOs. Were IXOs 17, 18 also QSOs,
the probability of their being background sources would be very low; however,
their optical spectra have still not been obtained, and  they will probably not be observed
in the near future because they are quite faint (they have no optical 
counterpart in the Palomar plates). For the two discovered QSOs (excluding 
IXOs 17, 18 from the list) the probabilities are: $P_X=3\times 10^{-3}$, 
$P_B=10^{-5}$. We suspect that there may be an error in the magnitudes
given by NOMAD for the optical counterpart of IXO 20 ($m_B=15.2$, $m_R=18.1$).

\item[NGC 3628]: with A9,10,13,14 as QSOs. Again, the most
intriguing case is IXO 39, with lower values of $\lambda $, but
this has not been observed spectroscopically yet since its optical counterpart
is not detected. With the remaining four QSOs (excluding IXO 39 from the
list): $P_X=2\times 10^{-5}$, $P_B=8\times 10^{-5}$. And there
are still another three QSOs in A04 which do not follow the constraint
$L_X[2-10 keV]\ge 10^{39}$ but which, if included, will considerably reduce 
the probability to three orders of magnitude. Precisely because of
that, it is a known  anomalous redshift candidate 
(Arp 2002), in which there is not only the overdensity but also an
alignment of QSOs across the nucleus of NGC 3628 and their connection 
with X-ray isophotes and coincidences in the alignment with some
other filaments in the radio and optical.

\item[NGC 4319]: with A21 as QSO. This known QSO, Mrk 205, is
indeed a well known controversial case in the debate about anomalous
redshifts. It is not only the low probability of finding so bright a QSO
near NGC 4319 ($P_X=4\times 10^{-6}$ and $P_B=10^{-6}$) but also the 
fact that it is
connected by a small bridge to the galaxy (Sulentic \& Arp 1987a,b). 
There is also an almost continuous luminous connection
extending from Mrk 205 into the nucleus of the spiral, and a corresponding
feature on the opposite side of the disc appearing to link a bright UV knot
($m_B=19.5$; with redshift of the galaxy, Sulentic \& Arp 1987b) 
with the nucleus (Sulentic \& Arp 1987a).

\item[NGC 4374]: with A22 as QSO. The probabilities alone for this object (excluding 
IXOs 50 and 51 from the list) are: $P_X=2\times 10^{-2}$, $P_B=3\times 10^{-3}$,
high enough for A22 most likely to be a background object.
IXOs 50 and 51 have not yet been observed. 
There is no optical counterpart in the Palomar plates; 
however there are
optical counterparts in SDSS-4th release survey: $u=24.63$, $g=25.11$, $r=24.80$
for IXO 50 (with a probable large error in these magnitudes because we are
at the limit of the detection) and $u=20.97$, $g=20.70$, $r=20.01$
for IXO 51. Both of them might be QSOs according to their colour [see eq.\ (1)
of L\'opez-Corredoira \& Guti\'errez 2004]. Were these two cases really
QSOs, we would have a very interesting new case of anomalous redshift.

\item[NGC 4472]: with IXOs 57, 58 as QSOs. And there remain IXOs 55, 56,
59, 60 and 61 to be identified. Only with the IXOs 57 and 58 as QSOs (excluding
the other five IXOs from the list), the probabilities that they are both
background objects are: $P_X=0.1$, $P_B=2\times 10^{-3}$. They are most
probably two background QSOs. 
IXOs 55 ($u=20.57$, $g=20.66$, $r=20.41$)
and 59 ($u=20.68$, $g=20.07$, $r=20.08$) have SDSS colours of QSOs 
[see eq.\ (1) of L\'opez-Corredoira \& Guti\'errez 2004], while IXOs 56
($u=21.53$, $g=20.71$, $r=20.59$), 60 ($u=22.89$, $g=21.36$, $r=20.81$), and
61 ($u=23.42$, $g=22.22$, $r=21.35$) are redder. 
Even where we had four QSOs (IXOs 55, 57, 58 and 59), the probability
would be high enough to be a fortuitous case of background sources.

\item[NGC 4649]: with IXOs 69 and 71 as QSOs. IXO 70 remains to be identified, 
although according to its SDSS colour ($u=19.34$, $g=19.32$, $r=19.09$), 
it might also be a QSO.
Only with the IXOs 69 as 71 as QSOs (excluding
IXO 70 from the list), the probabilities that they are both
background objects are: $P_X=10^{-2}$, $P_B=2\times 10^{-3}$. They are most
probably two background QSOs.

\end{description}

Therefore, the possibly controversial cases with regard to 
anomalous redshift are only those that were known in the past,
such as NGC 3628 and NGC 4319; the C02 catalog does not contain
clear cases of QSOs with different redshifts and low probabilities of 
being background objects.

Radecke (1997) points out a detection of correlation between
nearby Seyfert galaxies and X-ray sources, but the selection is
different from those used in the C02 catalogue. Radecke uses only Seyfert
galaxies while we use all kinds of galaxies. Radecke gets the maximum 
correlation for very bright sources with B magnitude lower than 10.5,
while most galaxies in C02 sample are fainter. Finally, Radecke puts
the constraint for X-ray sources having more than 5 counts/ks while
in the C02 sample most of the sources have less than 5 counts/ks.
Therefore, there is no inconsitency between Radecke's analysis (which
finds a number of X-ray sources around his galaxies considerably
higher than expected from background sources; in the best of the cases, he
gets excess over 250\%) and our result (we got an excess of $\sim 100$\% over
the background sources, which are due to ULXs belonging to the galaxies
but they are not QSOs).
It is also quite possible that part or all of the excess of
sources around Seyfert galaxies in Radecke's analysis is due to
the existence a considerable number of ULXs belonging to the galaxies,
these ULXs having the same redshift as the galaxies, so they could not
be used as a claim of anomalous redshift. However, this hypothesis
has not been tested yet.

\section{Conclusions}

From the above analysis we  conclude the following:

\begin{itemize}

\item More than a half of the ULX candidate in the C02c catalogue are
background QSOs. This result from the statistical
analysis is confirmed by follow-up spectroscopy of some of the sources.

\item The clean sample is as follows.
The  spectroscopically confirmed ULXs are
IXOs 5, 7, 8, 22, 31, 34, 65, 68, 72, 73, 77, 80, 81; 
plus IXOs 26, 27, 42, 62, 79, which were not observed spectroscopically
but  are bright diffuse sources, so they cannot be background QSOs
and are most probably HII regions of the parent galaxies.
Other objects that have not 
yet been observed spectroscopically,
and that have higher probabilities of being ULXs in the C02 catalogue are 
IXOs 17, 18, 25, 39, 66, 76 and 78.

\item Among the 87 sources in C02 catalogue, there are no significant
statistical cases that indicate non-cosmological redshifts in
some QSOs associated with galaxies. Nonetheless, the problem of the
anomalous redshifts is still open because there remain other
known cases, for instance those of NGC 4319 or NGC 3628, with significantly
low probabilities of their QSOs being background objects.

\end{itemize}

\

Acknowledgments: Thanks are given to the referee, P. Kaaret, for
helpful comments. The authors were supported by the {\it Ram\'on y
Cajal} Programme of the Spanish Science Ministry. Thanks are given
to T. J. Mahoney (IAC) for the proof-reading of the paper.

\begin{table*}
\caption{Column 1: RC3 galaxy. Column 2: number of IXO in
the C02 catalog, or from table 1 in A04 (named with Axx, and the number
xx of row in that table). Column 3: flux in the range from 0.5 to 2.0 keV
[units $10^{-14}$ erg/s/cm$^2$]. Column 4: average number of expected sources
up to the distance of this source to the centre of the galaxy, and
brighter than its flux in X-ray. Column 5: probability that all candidate 
ULXs associated to the galaxy are indeed background sources with
the observed distribution of X-ray fluxes.
Column 6: B magnitude of the most likely 
optical counterpart from NOMAD database; letter ``d'' indicates that
it is a diffuse source or a point-like source embedded in a halo;
letter ``a'' indicates that the source is in the spiral arm of the galaxy.
Column 7: average number of expected sources
up to the distance of this source to the centre of the galaxy and
brighter than its B magnitude. Column 8: probability that all candidate 
ULXs associated with the galaxy are indeed background sources with the
observed distribution of B magnitudes.
Column 9: spectroscopic identification [reference]. 
References of the spectroscopical identification: 1: A04; 
2: Zampieri et al. (2004); 3: Guti\'errez
\& L\'opez-Corredoira (2006); 4: Guti\'errez (2006); 
5: Lehmann et al. (2005);
6: Kaaret et al. (2004); 7: Guti\'errez \& L\'opez-Corredoira (2005);
8: Ramsey et al. (2005); 9: Liu et al. (2005a); 10: NED database;
11: Liu et al. (2004); 12: Arp (1977); 13: Wong et
al. (2005) [We call `QSO' those  objects classified by Wong et al.\ (2005) 
as AGN, since most of these sources are presumably point-like QSOs rather 
than extended galaxies]; 
14: Ghosh et al. (2006); 15: Pakull \& Mirioni (2002);
16: Mucciarelli et al. (2005); 17: Kaaret (2005).}
\label{Tab:main}
\begin{center} 
\begin{tabular}{lllllllll}

Galaxy & IXO  & $S_{X,i}$ &  $\lambda _{X,i}$  
&  $P_X$  & $m_{B,i}$ &  $\lambda _{B,i}$  & $P_B$ & spectr. ident. \\ \hline

NGC 720 & 1 & 1.1 & 0.97 & $5\times 10^{-2}$ &
20.1 & 0.16 & $7\times 10^{-4}$ & QSO [1,13] \\

       & 2 & 4.4 & $8.9\times 10^{-2}$ &  &
18.7 & $5.1\times 10^{-3}$ &  & QSO [1,13] \\

NGC 891 & 3 & 7.9 & 0.16 & 0.1 &
$>21.5$ & $>1.46$ & $>0.8$ &   \\

NGC 1042 & 4 & 17 & $3.5\times 10^{-3}$ & $3\times 10^{-3}$ &
$>21.5(a)$ & $>0.12$ & $>0.1$ &   \\

NGC 1073 & 5 & 2.2 & 0.12 & $10^{-4}$ &
13.1(a) & $1.2\times 10^{-7}$ & $8\times 10^{-12}$ & HII [17]  \\

         & A3 & 2.8 & $5.5\times 10^{-2}$ & &
V=20.0 & $2.4\times 10^{-2}$ &  & QSO [1]  \\

         & A5 & 7.0 & $2.7\times 10^{-2}$ & &
V=18.8 & $4.5\times 10^{-3}$ &  & QSO [1]  \\

NGC 1291 & 6 & 9.8 & $5.8\times 10^{-2}$ & $6\times 10^{-2}$ &
$>21.5$ & $>0.78$ & $>0.5$ &     \\

NGC 1313 & 7 & 210 & $1.5\times 10^{-5}$ & $6\times 10^{-8}$ &
$>21.5$ & $>3.2\times 10^{-2}$ & $>6\times 10^{-3}$ & HII [15]    \\

         & 8 & 84 & $3.8\times 10^{-3}$ &  &
20.0 & 0.47 &  & O-BV or GI in HII [2,8,15,16]    \\

NGC 1316 & 9 & 8.0 & 0.59 & $2\times 10^{-3}$ &
19.5 & 0.70 & $>3\times 10^{-4}$ &  QSO [13]   \\

         & 10 & 4.0 & 0.27 &    &
20.2 & 0.27 &  &   QSO [13]   \\

         & 11 & 2.5 & 0.36 &    &
$>21.5$ & $>0.51$ &  &     \\

         & 12 & 2.0 & 1.04 &    &
$>21.5$ & $>1.01$ &  &     \\

         & 13 & 2.0 & 2.95 &    &
21.2 & 2.33 &  &     \\

         & 14 & 2.5 & 3.05 &    &
19.1 & 0.23 &  &   QSO [13]   \\

NGC 1365 & 15 & 5.9 & 0.29 & $4\times 10^{-2}$ &
19.4 & 0.17 & $10^{-3}$ &   QSO [13]   \\

         & 16 & 5.9 & 1.01 &    &
18.7 & $2.7\times 10^{-2}$ &  &  QSO [13]    \\

         & A6 & 5.9 & 0.48 &    &
$V=19.7$ & 0.47 &  &  QSO [1]    \\

NGC 1399 & 17 & 3.1 & $2.9\times 10^{-2}$ & $10^{-7}$  &
$>21.5$ & $>5.7\times 10^{-2}$ & $>6\times 10^{-9}$ &     \\

         & 18 & 6.2 & $4.2\times 10^{-3}$ &    &
$>21.5$ & $>2.6\times 10^{-2}$ &  &     \\

         & 19 & 31 & $1.4\times 10^{-2}$ &    &
20.1 & 0.37 &  &   QSO [13]   \\

         & 20 & 4.9 & 0.23 &    &
15.2 & $4.3\times 10^{-5}$ &  &   QSO [13]   \\

NGC 1427A  & 21 & 5.0 & $4.3\times 10^{-3}$ & $4\times 10^{-3}$   &
18.4 & $2.9\times 10^{-4}$ & $3\times 10^{-4}$ &     \\

IC342  & 22 & 48 & $6.5\times 10^{-3}$ & $6\times 10^{-3}$   &
17.8 & $6.3\times 10^{-3}$ & $6\times 10^{-3}$ & HII [3,8,15] \\

NGC 1553  & 23 & 5.5 & 0.10 & 0.1   &
20.8 & 0.32 & 0.3 &     \\

NGC 1566  & 24 & 5.8 & $4.8\times 10^{-3}$ & $5\times 10^{-3}$   &
$>21.5(a)$ & $>2.7\times 10^{-2}$ & $>3\times 10^{-2}$ &     \\

NGC 1672  & 25 & 1.8 & 0.11 & $2\times 10^{-5}$   &
$>21.5(a)$ & $>9.8\times 10^{-2}$ & $\sim 2\times 10^{-14}$ &     \\

          & 26 & 5.6 & $1.1\times 10^{-2}$ &   &
15.1(da) & $2.9\times 10^{-4}$ &  &     \\

         & 27 & 4.5 & $2.5\times 10^{-2}$ &   &
R=13.4(da) & $\sim 1.4\times 10^{-7}$ &  &     \\

NGC 1792 & 28 & 3.4 & $1.3\times 10^{-2}$ & $10^{-2}$   &
$>21.5$ & $>3.0\times 10^{-2}$ & $>3\times 10^{-2}$ &     \\

NGC 1961  & 29 & 5.3 & $7.2\times 10^{-2}$ & $7\times 10^{-2}$   &
20.6 & 0.17 & 0.2 &    \\

Mrk 3  & 30 & 4.0 & $4.2\times 10^{-2}$ & $4\times 10^{-2}$   &
$>21.5$ & $0.13$ & $>0.1$ &     \\

Holmberg II  & 31 & 360 & $3.7\times 10^{-5}$ & $4\times 10^{-5}$   &
14.2 & $1.3\times 10^{-6}$ & $10^{-6}$ & O4V-B3Ib in HII [5,6,8,15]    \\

NGC 2775  & 32 & 2.2 & 0.55 & $6\times 10^{-2}$   &
18.1 & $5.7\times 10^{-3}$ & $2\times 10^{-4}$ & QSO [4,13]    \\

          & 33 & 4.4 & 0.16 &    &
19.2 & $3.9\times 10^{-2}$ &   &  QSO [7,13]    \\

NGC 3031  & 34 & 360 & $1.4\times 10^{-3}$ & $10^{-3}$  &
20.4 & 3.00 & 0.95  &  4-6 Myr blue stars in HII [8,15]   \\

NGC 3226  & 35 & 2.3 & 0.22 & 0.2  &
18.4 & $4.3\times 10^{-3}$ & $4\times 10^{-3}$  & local M star [4,13]    \\

NGC 3256  & 36 & 2.1 & 0.34 & 0.3  &
$>21.5$ & $>0.36$ & $>0.3$  &      \\

IC 2597  & 37 & 23 & $3.0\times 10^{-3}$ & $3\times 10^{-3}$  &
20.8 & $9.5\times 10^{-2}$ & 0.1  &  Emis.line Gal. [4,13]    \\

NGC 3310  & 38 & 4.1 & $1.6\times 10^{-2}$ & $2\times 10^{-2}$  &
17.5 & $1.5\times 10^{-4}$ & $10^{-4}$  &     \\

\end{tabular}
\end{center}
\end{table*}

\begin{table*}
Cont. table \protect{\ref{Tab:main}}.
\begin{center}
\begin{tabular}{lllllllll}
Galaxy & IXO  & $S_{X,i}$  &  $\lambda _{X,i}$  
&  $P_X$  & $m_{B,i}$ &  $\lambda _{B,i}$  & $P_B$ & spectr. ident. \\ \hline

NGC 3628  & 39 & 24 & $3.8\times 10^{-5}$ & $3\times 10^{-10}$  &
$>21.5$ & $>2.3\times 10^{-3}$ & $>8\times 10^{-8}$ &      \\

         & A9 & 770 & $9.0\times 10^{-2}$ &    &
$V=19.5$ & 0.10 &  &  QSO [1]    \\

         & A10 & 15 & $4.9\times 10^{-2}$ &    &
$V=19.2$ & $9.3\times 10^{-2}$ &  &  QSO [1]    \\

         & A13 & 9.7 & 0.22 &    &
$V=19.6$ & 0.42 &  &  QSO [1]    \\

         & A14 & 31 & $7.8\times 10^{-2}$ &    &
$V=18.3$ & $9.1\times 10^{-2}$ &  &  QSO [1]    \\

NGC 3923  & 40 & 7.3 & 0.11 & 0.1  &
19.0 & $3.7\times 10^{-2}$ & $4\times 10^{-2}$ &  QSO [4,13]    \\

UGC 7009  & 41 & 8.2 & $1.4\times 10^{-3}$ & $10^{-3}$  &
$>21.5$ & $1.4\times 10^{-2}$ & $10^{-2}$ &      \\

NGC 4088  & 42 & 5.0 & $2.7\times 10^{-3}$ & $3\times 10^{-3}$  &
14.8(da) & $2.4\times 10^{-7}$ & $3\times 10^{-7}$ &      \\

NGC 4151  & 43 & 2.2 & 1.02 & $6\times 10^{-4}$  &
18.8 & $2.6\times 10^{-2}$ & $8\times 10^{-4}$ & Emis.line Gal. [3,12,13]   \\

          & 44 & 1.1 & 0.46 &    &
20.9 & 0.16 &  &      \\

         & A15 & 62 & $3.8\times 10^{-3}$ &    &
$V=20.3$ & 0.41 &  &  QSO [1]    \\

NGC 4168 & A16 & 1.2 & $4.2\times 10^{-2}$ & $4\times 10^{-2}$  &
$V=18.7$ & $4.2\times 10^{-4}$ & $4\times 10^{-4}$ &  QSO [1]    \\

NGC 4203  & 45 & 2.2 & 0.29 & 0.3  &
18.3 & $4.3\times 10^{-3}$ & $4\times 10^{-3}$ & QSO [7]     \\

NGC 4254  & 46 & 14 & $6.2\times 10^{-3}$ & $6\times 10^{-3}$  &
$>21.5$ & $>0.15$ & $>0.1$ &      \\

NGC 4319  & A21 & 360 & $4.3\times 10^{-6}$ & $4\times 10^{-6}$  &
$V=15.2$ & $9.9\times 10^{-7}$ & $10^{-6}$ &  QSO [1]    \\

NGC 4321  & 47 & 3.2 & 0.81 & 0.2  &
20.3 & 0.64 & $>0.1$ &      \\

          & 48 & 1.6 & 2.41 &   &
$>21.5$ & $>1.92$ &  &      \\

          & 49 & 2.0 & 0.67 &   &
$>21.5$ & $>0.66$ &  &      \\

NGC 4374  & 50 & 16 & $2.0\times 10^{-4}$ & $2\times 10^{-7}$  &
$>21.5$ & $>6.3\times 10^{-3}$ & $>6\times 10^{-6}$ &      \\

          & 51 & 6.4 & 0.11 &   &
$>21.5$ & $>0.73$ &  &      \\

         & A22 & 10 & $1.8\times 10^{-2}$ &    &
$V=18.5$ & $3.1\times 10^{-3}$ &  &  QSO [1]    \\

%

NGC 4373  & 52 & 2.2 & 0.20 & 0.2  &
18.7 & $3.7\times 10^{-3}$ & $4\times 10^{-3}$ &  QSO [13]    \\

NGC 4395  & 53 & 35 & $3.1\times 10^{-3}$ & $3\times 10^{-3}$  &
$>21.5$ & $>0.36$ & $>0.3$ &      \\

NGC 4438  & 54 & 13 & $2.6\times 10^{-2}$ & $3\times 10^{-2}$  &
19.5 & $6.9\times 10^{-2}$ & $7\times 10^{-2}$ & QSO [3]    \\

NGC 4472  & 55 & 2.0 & 3.44 & $10^{-4}$  &
20.8 & 2.03 & $>3\times 10^{-5}$ &      \\

          & 56 & 4.1 & 1.14 &   &
20.8 & 2.13 &  &      \\

          & 57 & 10 & 0.21 &   &
18.2 & $3.4\times 10^{-2}$ &  & QSO [3]     \\

          & 58 & 4.1 & 0.96 &   &
18.4 & $4.7\times 10^{-2}$ &  & QSO [7]    \\

          & 59 & 3.2 & 2.19 &   &
21.0 & 3.27 &  &      \\

          & 60 & 13 & $9.6\times 10^{-2}$ &   &
21.0 & 1.42 &  &      \\

          & 61 & 2.6 & 0.93 &   &
$>21.5$ & $>1.36$ &  &      \\

NGC 4485  & 62 & 11 & $2.7\times 10^{-4}$ & $3\times 10^{-4}$  &
13.2(d) & $4.4\times 10^{-9}$ & $4\times 10^{-9}$ &      \\

NGC 4552  & 63 & 5.1 & 0.26 & $5\times 10^{-2}$  &
20.4 & 0.48 & $5\times 10^{-2}$ &      \\

          & 64 & 3.2 & 0.25 &   &
20.0 & 0.14 &  &      \\

NGC 4559  & 65 & 39 & $1.5\times 10^{-3}$ & $5\times 10^{-8}$  &
12.6 & $7.0\times 10^{-8}$ & $>10^{-10}$ &  4 blue/red superg. in HII [3,9,15]    \\

          & 66 & 24 & $3.4\times 10^{-5}$ &  &
$>21.5$ & $>2.1\times 10^{-3}$ &  &      \\

NGC 4565  & 67 & 21 & $5.0\times 10^{-4}$ & $5\times 10^{-4}$ &
$>21.5$ & $>2.5\times 10^{-2}$ & $>2\times 10^{-2}$ &      \\

NGC 4631  & 68 & 12 & $1.6\times 10^{-2}$ & $2\times 10^{-2}$ &
V=14.2(da) & $\sim 2.1\times 10^{-6}$ & $\sim 2\times 10^{-6}$ & HII [15]  \\

NGC 4649  & 69 & 5.2 & $8.8\times 10^{-2}$ & $4\times 10^{-3}$ &
18.3 & $5.4\times 10^{-3}$ & $4\times 10^{-5}$ &  QSO [7]    \\

          & 70 & 4.1 & 0.71 &  &
18.4 & $3.6\times 10^{-2}$ &  &     \\

          & 71 & 10 & 0.16 &  &
19.8 & 0.46 &  & QSO [7]     \\

NGC 4698 & A23 & 10 & $4.8\times 10^{-3}$ & $5\times 10^{-3}$  &
$V=20.5$ & $3.0\times 10^{-2}$ & $3\times 10^{-2}$ &  QSO [1]    \\

NGC 4861  & 72 & 9.1 & $4.9\times 10^{-3}$ & $9\times 10^{-7}$  &
9.5(a) & $6.9\times 10^{-11}$ & $>10^{-12}$ & HII [10,15]     \\

          & 73 & 29 & $1.9\times 10^{-4}$ &  &
$>21.5$ & $>1.6\times 10^{-2}$ &  &  HII [15]    \\

NGC 5055  & 74 & 35 & $1.3\times 10^{-2}$ & $10^{-2}$ &
18.4 & $2.3\times 10^{-2}$ & $2\times 10^{-2}$ &      \\

NGC 5128  & 75 & 38 & $9.2\times 10^{-3}$ & $3\times 10^{-6}$  &
10.7 & $1.3\times 10^{-8}$ & $>4\times 10^{-9}$  &  local M star [10,13]   \\

          & 76 & 150 & $2.9\times 10^{-4}$ &  &
$>21.5$ & $>0.36$ &  &      \\

NGC 5204  & 77 & 63 & $1.3\times 10^{-5}$ & $10^{-5}$ &
$>21.5$ & $>3.8\times 10^{-3}$ & $>4\times 10^{-3}$ &  B0Ib in nebula [11,15]  \\

NGC 5194  & 78 & 6.5 & $3.8\times 10^{-2}$ & $2\times 10^{-8}$ &
$>21.5(a)$ & $>0.26$ & $>5\times 10^{-13}$ &     \\

          & 79 & 81 & $1.3\times 10^{-2}$ &  &
13.0(da) & $9.8\times 10^{-8}$ &  &   \\

          & 80 & 16 & $8.1\times 10^{-3}$ &  &
16.9(a) & $2.5\times 10^{-4}$ & & young star cluster [9]   \\

          & 81 & 13 & $1.4\times 10^{-2}$ &  &
$>21.5$(a) & $>0.29$ &  & ? in young region [9]  \\

NGC 5236  & 82 & 33 & $9.8\times 10^{-3}$ & $10^{-2}$ &
$>21.5$ & $>0.98$ & $>0.6$ &   \\

NGC 5457  & 83 & 20 & 0.12 & 0.1 &
$>21.5$ & $>4.99$ & $>0.99$ &   \\

NGC 5775  & 84 & 2.3 & 0.43 & 0.3 &
19.6 & $7.6\times 10^{-2}$ & $7\times 10^{-2}$ &  QSO [7,14] \\

NGC 6946  & 85 & 15 & $1.9\times 10^{-3}$ & $2\times 10^{-3}$ &
$>21.5$(a) & $>5.2\times 10^{-2}$ & $>5\times 10^{-2}$ &   \\

NGC 7314  & 86 & 3.2 & $6.4\times 10^{-2}$ & $6\times 10^{-2}$ &
18.6 & $1.7\times 10^{-3}$ & $>2\times 10^{-3}$ &   \\

NGC 7590  & 87 & 6.4 & $1.3\times 10^{-3}$ & $10^{-3}$ &
$>21.5$(a) & $>8.8\times 10^{-3}$ & $>9\times 10^{-3}$ &   \\

\end{tabular}
\end{center}
\end{table*}


\begin{thebibliography}{99}

\bibitem{} Adelman-McCarthy, J. K., Ag\"ueros, M. A., Allam, S. S.,
et al., 2006, ApJS 162, 38

\bibitem{} Arp, H. C., 1977, ApJ 218, 72

\bibitem{} Arp, H. C., 2002, A\&A 391, 833

\bibitem{} Arp, H. C., Guti\'errez, C. M., L\'opez-Corredoira, M.,
2004, A\&A, 418, 877 (A04)

\bibitem{} Arp, H. C., \& Sulentic, J. W. 1979, ApJ 229, 496

\bibitem{} Boyle, B. J., Shanks, T., Croom, S. M., et al., 2000,
MNRAS, 317, 1014

\bibitem{} Burbidge G., Burbidge E. M., \& Arp H., 2003, A\&A 400, L17

\bibitem{} Cheney, J. E., \& Rowan-Robinson, M., 1981, MNRAS, 195, 497

\bibitem{} Clark, D. M., Christopher, M. H., Eikenberry, S. S., et al.,
2005, ApJ 631, L109

\bibitem{} Colbert, E. J. M., \& Mushotzky, R. F., 1999, ApJ,
519, 89

\bibitem{} Colbert, E. J. M., \& Ptak, A. F., 2002, ApJS, 143, 25 (C02)

\bibitem{} Gao, Y., Wang, Q. D., Appleton, P. N., \& Lucas, R. A.,
2003, ApJ 596, L171

\bibitem{} Ghosh, K. K., Finger, M. K., Swartz, D. A.,
Tennannt, A. F., \& Wu, K., 2005, astro-ph/0509720

\bibitem{} Ghosh, K. K., Swartz, D. A., Tennant, A. F., Saripalli, L.,
Gandhi, P., Foellmi, C., Guti\'errez, C. M., \& L\'opez-Corredoira,
M., 2006, in preparation

\bibitem{} Guti\'errez, C. M., 2006, ApJ-Lett., accepted;
astro-ph/0602207

\bibitem{} Guti\'errez, C. M., L\'opez-Corredoira, M., 2005,
ApJ, 622, L89

\bibitem{} Guti\'errez, C. M., L\'opez-Corredoira, M., 2006,
in preparation

\bibitem{} Hao, L., Strauss, M. A., Fan, X., et al. 2005, AJ, 129, 1795

\bibitem{} Hasinger, G., 1998, Astron. Nachr., 319, 37

\bibitem{} Hasinger, G., Burg, R., Giacconi, R., Hartner, G.,
Schmidt, M., Tr\"umper, J., \& Zamorani, G., 1993, A\&A, 275, 1

\bibitem{} Irwin, J. A., Bregman, J. N., \& Athey, A. E., 2004,
ApJ 601, L143

\bibitem{} Kaaret, P., 2005, ApJ 629, 233

\bibitem{} Kaaret, P., Ward, M.-J., \& Zezas, A. 2004, MNRAS 351, L83

\bibitem{} Kilgard, R. E., Kaaret, P., Krauss, M. I., 
Prestwich, A. H., Raley, M. T., \& Zezas, A., 2002, ApJ, 573, 138

\bibitem{} Larsen J. A., Humphreys R. M. 2003, AJ 125, 1958

\bibitem{} Lehmann, I., Becker, T., Fabrika, S., et al. 2005,
A\&A 431, 847

\bibitem{} Lira, P., Lawrence, A., \& Johnson, R. A., 2000,
MNRAS, 319, 17

\bibitem{} Liu, J.-F.. Bregman, J.-N., 2005, ApJS 157, 59

\bibitem{} Liu, J.-F., Bregman, J.-N., \& Seitzer, P., 2004, ApJ
602, 249

\bibitem{} Liu, J.-F., Bregman, J.-N., Seitzer, P., \&
Irwin, J., 2005a, astro-ph/0501310

\bibitem{} Liu, J.-F., Bregman, J. N., \& Irwin, J., 2005b,
astro-ph/0501312

\bibitem{} Liu, Q.-Z., \& Mirabel, I. F., 2005, A\&A 429, 1125

\bibitem{} L\'opez-Corredoira, M., \& Guti\'errez, C. M., 2004,
A\&A, 421, 407

\bibitem{} Makishima, K., Kubota, A., Mizuno, T., et al., 2000,
ApJ 535, 632

\bibitem{} Meyer, M. J., Drinkwater, M. J., Phillips, S., \& Couch,
W. J., 2001, MNRAS, 324, 343

\bibitem{} Mucciarelli, P., Zampieri, L., Falomo, R.,
Turolla, R., \& Treves, A., 2005, ApJ 633, L101

\bibitem{} Pakull, M. W., \& Mirioni, L., in:
New Visions of the X-ray Universe in the XMM-Newton and Chandra Era, 
ESTEC, The Netherlands. Preprint astro-ph/0202488

\bibitem{} Ptak, A., \& Colbert, E., 2004, ApJ 606, 291

\bibitem{} Radecke, H. D., 1997, A\&A, 319, 18

\bibitem{} Ramsey, C. J., Williams, R. M., Gruendl, R. A.,
Chen, C.-H. R., Chu, Y.-H., \& Wang, Q. D., 2005, astro-ph/0511540

\bibitem{} Robin, A. C., Reyl\'e, C., Derri\`ere, S., \& Picaud, S.,
2003, A\&A, 409, 523

\bibitem{} Smith, B. J., Struck, C., Nowak, M. A. 2005, AJ 129, 1350

\bibitem{} Sulentic, J. W., \& Arp, H. C. 1987a, ApJ 319, 687

\bibitem{} Sulentic, J. W., \& Arp, H. C. 1987b, ApJ 319, 693

\bibitem{} Swartz, D., Ghosh, K., Tennant, A., \& Wu, K.,
2004, ApJS 154, 519

\bibitem{} Voges, W., et al. 1999, in Diffuse thermal and relativistic 
plasma in galaxy clusters, ed. H. Bohringer, L. Feretti, \& P. Schuecker 
(Garching: Max-Planck-Institut für Extraterrestrische Physik), 179

\bibitem{} Wong D. S., Chornock R., \& Filippenko A. V. 2005,
\textit{Populations of High Energy Sources in Galaxies, Proceedings of IAU
Symposium No. 230, Eds. Evert J.A. Meurs \& Giuseppina Fabbiano}, in press 

\bibitem{} Zampieri, L., Mucciarelli, P., Falomo, R., Kaaret, P., 
Di Stefano, R., Turolla, R., Chieregato, M., \& Treves, A., 2004, ApJ, 603, 523

\bibitem{} Zickgraf, F.-J., Engels, D., Hagen, H.-J., Reimers, D.,
\& Voges, W., 2003, A\&A 406, 535

\end{thebibliography}
\end{document}